\documentclass[prd,showpacs,preprintnumbers,nofootinbib,aps]{revtex4}
\usepackage{subeqnarray}
\usepackage{epsfig,amsmath,amssymb}
\usepackage{mathrsfs}
\usepackage[usenames,dvipsnames]{color}
\usepackage[pagebackref=false, colorlinks=false]{hyperref}
\definecolor{redish}{rgb}{0.7,0.2,0.0}  
\definecolor{bluish}{rgb}{0.2,0.5,0.8}
\hypersetup{linkcolor=redish,          
                  citecolor=blue,        
                  filecolor=magenta,      
                  urlcolor=bluish}          
\DeclareFontFamily{U}{rsfs}{}         
\DeclareFontShape{U}{rsfs}{m}{n}{<5> rsfs5 <6><7> rsfs7          %
  <8><9><10><10.95><12><14.4><17.28><20.74><24.88> rsfs10}{}     %
\DeclareMathAlphabet{\mathfs}{U}{rsfs}{m}{n}                     %

\newcommand{\ba}{\nopagebreak[3]\begin{eqnarray}}
\newcommand{\ea}{\end{eqnarray}}
\newcommand{\bii}{\begin{itemize}}
\newcommand{\eii}{\end{itemize}}

\begin{document}

\title{Geometric Phase using Diagonal Coherent State Represenatation}
\author{Prosenjit Maity}
\email{pmaityrkmrc08@gmail.com}
\author{Sobhan Sounda}
\email{sounda6@gmail.com}
\affiliation{Vivekananda Centre for Research (R.K.Mission Residential College), Kolkata-700103, India}
 \vspace{3 cm}      
\begin{abstract}

A given density operator of a two mode optical beam is evolved in complex projective ray space, first to another density operator corresponding to a specific angle of polarization, and then to a third density operator of different polarization angle. Calculating the Bargmann invariant corresponding to the trace of the product of these three density operators, its argument is shown to yield the geometric phase represented in terms of phase space variables and having the connotation of symplectic area. Our approach uses the Sudarshan's diagonal representation evading issues associated with the Fock basis.
\end{abstract}
\maketitle

\section{Introduction}

The concept of geometric phase was introduced in 1956 by S. Pancharatnam[1] in the context of polarization optics. In 1984 M.V. Berry [2] first realised that the geometric phase is also a feature of quantum mechanics. Later in 1993, N. Mukunda and R. Simon[3] developed a more general approach towards the quantum theory of geometric phase where the non-transitivity of Bargmann invariant plays the central role. In our present work we have shown that using this kinematic approach geometric phase can be derived by representing the density operator in diagonal coherent state basis instead of fock  state basis. The coherent state basis has a profound application in quantum optics. The salient features of these states are that they are linearly dependent of each other (mutually non-orthogonal) and form an over-complete set of basis leads to the famous  diagonal representation of any self adjoint operator due to E.C.G. Sudarshan[4]. The usefulness of our approach is that one can leisurely  avoid the messier part of the calculation which arises if one continues to evolve the density operator in the fock state basis. \\[.5mm]
For a given density operator  $\hat{\rho_1}$ ̂(which is on $\cal R$) for the two mode non-classical optical beam, we evolve it in  $\cal R$ to $\hat{\rho}_2$  and to  $\hat{\rho}_3$  by unitary operator $\mathrm{\hat{U}_p(\theta)}$. The curves connecting between  $\hat{\rho}_1$ to  $\hat{\rho}_2$ ,  $\hat{\rho}_2$ to $\hat{\rho_3}$ and  $\hat{\rho_3}$ to $\hat{\rho}_1$ are Null phase curves (NPC)[5,6]. In this evolution we change the state of polarization of the beam by using   unitary operator $\mathrm{\hat{U}_p}$ which is 
built by introducing the Jordan mapping of the generator of polarization for an arbitrary polariser into an operator which is a function of annihilation and creation operators. Construction of unitary operator for a polariser is given in section II. In section III we express $\hat{\rho}_1$ , $\hat{\rho}_2$   and   $\hat{\rho}_3$ in the diagonal coherent state basis which was first discovered in its entirety by E.C.G Sudarshan [4] and acknowledged its usefulness by R.J.Glauber[7]. In section IV the calculation of quasi-probability distribution function for a given density operator using anti-diagonal method introduced by C.L.Mehta[8] is shown. In Sudarshan's diagonal representation of density operator, the quasi-probability distribution function plays a significant role as it contains maximum amount of information of the state of the physical system. It's not a true probability distribution function[9] in the sense that it can  also be negative contradicts with the classical probability distribution. In section V calculation of geometric phase is illustrated. Concluding remarks are given in section VI.

\vspace{.3 in}
\section{Unitary operator corresponding to a polariser}

 we calculate the geometric phases for pure quantum mechanical state for which the density operator satisfies the 
properties $$\hat{\rho}^\dagger=\hat{\rho},\,\hat{\rho}^2=\hat{\rho},\, Tr(\hat{\rho}^2)=1$$\
The transmission matrix for the polariser $\mathrm{T_p}=\begin{pmatrix}
\mathrm{cos^2\theta} & \mathrm{cos\theta\, sin\theta}\\
\mathrm{sin\theta\, cos\theta} & \mathrm{sin^2\theta}
\end{pmatrix}$\\[5mm]
Differentiating the elements in $\mathrm{T_{p}}$  with respect to $\theta$ and evaluating it at $\theta$ equals to 
zero, we obtain the generator. Consider a plane monochromatic beam of light with two orthogonal polarisation 
modes. The annihilation and creation operator for each mode can be represented  as \,$\hat{a}^\dagger_j$,\, $\hat{a}_j;(j=1,2)$ respectively and we define \\[1mm]  
\hspace*{3cm}$\hat{a}^\dagger\equiv\begin{pmatrix}
{\hat{a}_1}^\dagger & {\hat{a}_2}^\dagger
\end{pmatrix}$ \hspace{1.5cm}and \hspace*{1cm} $\hat{a}=\begin{pmatrix}
\hat{a}_1\\
\hat{a}_2
\end{pmatrix}$\\
Introducing Jordan mapping we can write the unitary operator for polariser as $$\mathrm{\hat{U}_p(\theta)=exp\{i\theta({\hat{a}_1}^\dagger 
\hat{a}_2+{\hat{a}_2}^\dagger \hat{a}_1)\}}$$
The action of the polariser on annihilation operator can be described as\\[1mm]\hspace*{3cm}$\begin{pmatrix}
\hat{a}^\prime_1\\
\hat{a}^\prime_2
\end{pmatrix}$\,=\,$\mathrm{\begin{pmatrix}
cos\theta & isin\theta\\
isin\theta & cos\theta
\end{pmatrix}}$ $\begin{pmatrix}
\hat{a}_1\\
\hat{a}_2
\end{pmatrix}$ \hspace{.3cm}  and similarly acts on $\begin{pmatrix}
\hat{a}^\dagger_1\\
\hat{a}^\dagger_2
\end{pmatrix}$
 \vspace{1 in}

\section{Glauber-Sudarshan's P-representation of density operator and its unitary evolution}

The density operator $\hat{\rho}_1$ characterised by the states of a quantised electromagnetic beam with two mode can be 
written in P-representation [9, 10] as \\
\begin{equation}
\mathrm{\hat{\rho}_1=\,\frac{1}{\pi^2}\int\int d^2z_1d^2z_2\, P(z_1,z_2)\vert z_1,z_2\rangle \langle z_1,z_2\vert}
\end{equation}

\hspace{1cm}where $\vert{z_1,z_2}\rangle$ is two mode coherent state.
This initial statistical beam passes through a polariser which changes  $\hat{\rho}_1$  unitarily to\, $\hat{\rho}_2$. Hence 
$$\mathrm{\hat{\rho}_2={\hat{U}}^\dagger_p(\theta_1) \hat{\rho}_1\hat{U}_p(\theta_1)= \,\frac{1}{\pi^2}\int\int d^2z_1d^2z_2\, 
P(z_1,z_2){\hat{U}^\dagger}_p\vert z_1,z_2\rangle \langle z_1,z_2\vert {\hat{U}_p}}$$\
Since $\vert z_j\rangle$ can be obtained by operating the displacement operator $\mathrm{\hat{D}(z_j)=exp\{z_j\hat{a}^\dagger_j-z_j^\ast \hat{a}_j\}}$  
on vacuum state,
\begin{equation}
\mathrm{\hat{\rho}_2=\,\frac{1}{\pi^2}\int\int d^2z^\prime_1d^2z^\prime_2\, P({z^\prime}_1,{z^\prime}_2)\vert 
{z^\prime}_1,{z^\prime}_2\rangle \langle {z^\prime}_1,{z^\prime}_2\vert}
\end{equation}

\hspace{1cm}where $\mathrm{z^\prime_1= z_1\,cos\theta_1-iz_2 \,sin\theta_1}$; \, $\mathrm{z^\prime_2= 
z_2\,cos\theta_1-iz_1 \,sin\theta_1}$\\[2mm]
Similarly in the next evolution we set the polariser at an angle $\theta_2$ such that\\
\begin{equation}
\mathrm{\hat{\rho}_3=\,\frac{1}{\pi^2}\int\int d^2z^{\prime\prime}_1d^2z^{\prime\prime}_2\, 
P({z^{\prime\prime}}_1,{z^{\prime\prime}}_2)\vert
{z^{\prime\prime}}_1,{z^{\prime\prime}}_2\rangle \langle {z^{\prime\prime}}_1,{z^{\prime\prime}}_2\vert}
\end{equation}\\
Now the trace of these three density operators is \\[1mm]
$\mathrm{Tr(\hat{\rho}_1 \hat{\rho}_2 \hat{\rho}_3)}$\\
\hspace*{1cm} $$=\mathrm{\frac{1}{\pi^6}\int\int d^2 z_1\,d^2 z_2 P(z_1,z_2) \int\int d^2z^\prime_1d^2z^\prime_2 
P({z^\prime}_1,{z^\prime}_2\int\int d^2z^{\prime\prime}_1d^2z^{\prime\prime}_2 
P({z^{\prime\prime}}_1,{z^{\prime\prime}}_2)}$$\\
\,\hspace*{3cm}$\times \langle z_1,z_2\vert z^\prime_1,z^\prime_2\rangle \,\langle z^\prime_1,z^\prime_2\vert{z}^{\prime\prime}_1,{z}^{\prime\prime}_2\rangle 
\,\langle {z}^{\prime\prime}_1,{z}^{\prime\prime}_2\vert z_1,z_2\rangle$\\[.5cm]
Since $\mathrm{\langle z_1,z_2\vert z^\prime_1,z^\prime_2\rangle}$\\
\,$\displaystyle{\mathrm{= 
exp\{-\frac{1}{2}(|z_1|^2+|z_2|^2+|z^\prime_1|^2+|z^\prime_1|^2)\}\sum\limits_{n_1,n_2,n^\prime_1,n^\prime_2=0}^{\infty}\frac{(z^\ast_1)^{n_1}(z^\ast_2)^{n_2}(z^\prime_1)^{n_1} 
(z^\prime_2)^{n_2}}{\sqrt{n_1!\, n_2!\, n^\prime_1!\, n^\prime_2!}} \delta_{n_1,n^\prime_1}\,\delta_{n_2, 
n^\prime_2}}}$\\[1cm]
\,$\mathrm{=exp\{-\frac{1}{2}(|z_1|^2+|z_2|^2+|z^\prime_1|^2+|z^\prime_1|^2)+z^\ast_1 z^\prime_1+z^\ast_2 
z^\prime_2\}}$\\[1cm]
Therefore\, $\mathrm{Tr(\hat{\rho}_1\,\hat{\rho}_2\,\hat{\rho}_3)}$

$$\displaystyle{=\mathrm{\frac{1}{\pi^6}\int\int d^2 z_1\,d^2 z_2 P(z_1,z_2) \int\int d^2z^\prime_1d^2z^\prime_2 
P({z^\prime}_1,{z^\prime}_2)\int\int d^2z^{\prime\prime}_1 d^2z^{\prime\prime}_2 
P({z^{\prime\prime}}_1,{z^{\prime\prime}}_2)}}\\$$
$$\,\hspace*{2.5cm}\mathrm{\times\,\hspace{1mm} 
exp\{-(|z_1|^2+|z_2|^2+|z^\prime_1|^2+|z^\prime_1|^2)+|{z}^{\prime\prime}_1|^2+ |{z}^{\prime\prime}_2|^2)\}}$$
\begin{equation}
\hspace*{2.5cm}\mathrm{\times\,\hspace*{1mm} exp\{z^\ast_1 z^\prime_1+z^\ast_2 z^\prime_2+{z}^{\prime\ast}_1 
{z}^{\prime\prime}_1+{z}^{\prime\ast}_2 {z}^{\prime\prime}_2+{z}^{\prime\prime\ast}_1 {z}^{\prime\prime}_1 
+{z}^{\prime\prime\ast}_2 {z}^{\prime\prime}_2\}}
\end{equation} 
\vspace{1cm}

\section{Calculation of quasi-probability distribution function $\mathrm{P(z_1,z_2)}$  for two mode electromagnetic beam}
For two mode electromagnetic beam the density operator $\rho$\, can be written in diagonal form using coherent state 
basis as $$\mathrm{\hat{\rho} =\,\frac{1}{\pi^2}\int\int d^2z_1d^2z_2\, P(z_1,z_2)\vert z_1,z_2\rangle \langle z_1,z_2\vert}$$\
Let  us multiply both sides of the equation by the coherent state  $\mathrm{ \langle -v_1,-v_2\vert}$  on the left side and by 
coherent state $\mathrm{\vert v_1,v_2\rangle}$  on the right, where  $\mathrm{v_1}$  and $\mathrm{v_2}$ are some complex 
number. This anti-diagonal approach is due to C.L. Mehta[13]. We then obtain,\\[.1cm]

$\mathrm{\langle -v_1,-v_2\vert \hat{\rho}\vert  v_1,v_2\rangle}$\\
$$\mathrm{\,\hspace{1cm} = \frac{1}{\pi^2}\int\int d^2z_1d^2z_2\, P(z_1,z_2) \langle-v_1,v_2\vert z_1,z_2 \rangle \langle z_1,z_2\vert v_1,v_2\rangle}$$\\
$$\displaystyle{\mathrm{= \,\frac{1}{\pi^2}\, e^{-(\vert v_1\vert^2+\vert v_2 \vert^2)}\int\int d^2z_1d^2z_2\, P(z_1,z_2)e^{-(\vert z_1\vert^2+\vert z_2 \vert^2)}\, e^{v_1 z^\ast_1-v^\ast_1z_1+v_2 z^\ast_2-v^\ast_2 z_2}\, d^2z_1\, d^2z_2}} $$\\
We invert the Fourier integral in the above equation and obtain\\ [1 cm]
$\mathrm{P(z_1,z_2\,e^{-(\vert z_1\vert^2+\vert z_2\vert^2)})}$\\
\begin{equation}
\hspace{.5cm}\displaystyle{\mathrm{= \,\frac{1}{\pi^6} \int\int e^{\vert v_1\vert^2+\vert v_2 \vert^2}\,\langle -v_1,-v_2\vert \hat{\rho}\vert  v_1,v_2\rangle e^{-v_1 z^\ast_1+v^\ast_1z_1-v_2 z^\ast_2+v^\ast_2 z_2}\,d^2 v_1\, d^2 v_2 }}
\end{equation}
\newline
Let consider two mode electromagnetic beam described by the density operator $\hat{\rho}=\vert 1,1\rangle\langle 1,1\vert$. Obviously it is hermitian and corresponds to pure state. Now the quasi-probability distribution function can be obtained using eqn.(5) as 
$$\mathrm{P(z_1,z_2)=  \,\frac{1}{\pi^6}\, e^{(\vert z_1\vert^2+\vert z_2 \vert^2) }\, \int \vert v_1\vert^2 e^{-v_1 z^\ast_1+v^\ast_1z_1}\, d^2v_1\, \int \vert v_2\vert^2 e^{-v_2 z^\ast_2+v^\ast_2 z_2}\, d^2v_2}$$\\[.3cm]
Introducing\, $\mathrm{z=q+ip}$ where $\mathrm{(q, p) \in \mathbb{R}}$\, we obtain the probability distribution function in terms of phase space variables as\\
$$\mathrm{P(q_1,p_1;\,q_2,p_2)=\,\frac{1}{(4\pi^2)^2}\, e^{(q^2_1+p^2_1+q^2_2+p^2_2)}\,\{\delta(q_1)\,\delta^{(2)}(p_1)+\, \delta^{(2)}\,(q_1)\delta(p_1)\}}$$  
\begin{equation}
\mathrm{\hspace{3cm}\times\{\delta(q_2)\,\delta^{(2)}(p_2)+\, \delta^{(2)}\,(q_2)\,\delta(p_2)\}}
\end{equation}\\
where $\mathrm{\delta^{(2)}(q)}$ represents the derivative of delta function with respect to its argument. Now we shift the singular points  of delta function to some arbitrary points in phase space say $\mathrm{(q_{01},p_{01};q_{02},p_{02})}$ without loss of any generality. The modified distribution function is \\[.5cm]
 $\mathrm{P(q_1,p_1;\,q_2,p_2)}$\\
 $$\displaystyle{\mathrm{ =\,\frac{1}{(4\pi^2)^2}\, e^{(q^2_1+p^2_1+q^2_2+p^2_2)}\,\{\delta(q_1-q_{01})\,\delta^{(2)}(p_1-p_{01})+\, \delta^{(2)}\,(q_1-q_{01})\delta(p_1-p_{01})\}}}$$  \hspace{1.5cm}$\displaystyle{\mathrm{\hspace{3cm}\times\{\delta(q_2-q_{02})\,\delta^{(2)}(p_2-p_{02})+\, \delta^{(2)}\,(q_2-q_{02})\,\delta(p_2-p_{02})\}}}$\\[.8cm]
 $ \mathrm{\hspace{2cm}=\,P(q_1,p_1)\,P(q_2,p_2)}$\\[1cm]
 This is obvious because the two mode of electromagnetic beam are independent of each other and there is no coupling between them.\\[3 in]
 
 \section{Calculation of geometric phase}
 
Rewriting eqn.(4) in terms of phase space variables we obtain \\[.5cm]
 $\mathrm{Tr(\hat{\rho}_1\hat{\rho}_2\hat{\rho}_3)}$\\
 $$\mathrm{= \frac{1}{\pi^6}\,\int\int e^{-(q^2_1+p^2_1)}\, P(q_1,p_1)\,d^2 q_1\,d^2p_1\, \int\int e^{-(q^2_2+p^2_2)}\, P(q_2,p_2)\,d^2 q_2\,d^2p_2} $$ 
$$\displaystyle{\mathrm{\times\int\int e^{\{q_1 q^{\prime}_1 + p_1 p^{\prime}_1 -({q^\prime}^2_1+{p^\prime}^2_1)+i(q_1 p^{\prime}_1- p_1 q^{\prime}_1)\}}\, P^\prime(q^{\prime}_1, p^{\prime}_1)\,d^2 q^{\prime}_1\,d^2 p^{\prime}_1}}$$
$$\displaystyle{\mathrm{\times\int\int e^{\{q_2 q^{\prime}_2 + p_2 p^{\prime}_2 -({q^{\prime}}^2_2+{p^{\prime}}^2_2)+i(q_2 p^{\prime}_2- p_2 q^{\prime}_2)\}}\, P^\prime(q^{\prime}_2, p^{\prime}_2)\,d^2 q^{\prime}_2\,d^2 p^{\prime}_2}}$$
$$\displaystyle{\mathrm{\times \int\int e^{\{{q^\prime}_1 {q}^{\prime\prime}_1+{p^\prime}_1 {p}^{\prime\prime}_1 +{q}^{\prime\prime}_1 q_1+{p}^{\prime\prime}_1 p_1-({q}^{{\prime\prime}^2}_1+{p}^{{\prime\prime}^2}_1 )+i{(q^\prime_1 {p}^{\prime\prime}_1 - {q}^{\prime\prime}_1 p^\prime_1+{q}^{\prime\prime}_1 p_1-q_1 {p}^{\prime\prime}_1)}\}}\,P^{\prime\prime}(q^{\prime\prime}_1, p^{\prime\prime}_1)\,d^2q^{\prime\prime}_1\,d^2p^{\prime\prime}_1}} $$
\begin{equation}
\displaystyle{\mathrm{\times \int\int e^{\{{q^\prime}_2 {q}^{\prime\prime}_2 +{p^\prime}_2 {p}^{\prime\prime}_2 +{q}^{\prime\prime}_2 q_2 + {p}^{\prime\prime}_2 p_2 -({q}^{{\prime\prime}^2}_2+ {p}^{{\prime\prime}^2}_2 )+i{(q^\prime_2 {p}^{\prime\prime}_2 - {q}^{\prime\prime}_2 p^\prime_2+{q}^{\prime\prime}_ 2 p_2- q_2 {p}^{\prime\prime}_2)}\}}\,P^{\prime\prime}(q^{\prime\prime}_2, p^{\prime\prime}_2)\,d^2q^{\prime\prime}_2\,d^2p^{\prime\prime}_2}} 
\end{equation}\\

Here $\mathrm{P^\prime (q^\prime_1,p^\prime_1)}$ and $\mathrm{P^{\prime\prime}(q^{\prime\prime}_1, p^{\prime\prime}_1)}$ are both combination of the Dirac delta function and its second order derivatives because unitary evolution doesn't affect the distribution function rather it changes the state vectors and hence the variables  associated with the state vectors.\\[.5cm]
After performing the integration and taking the argument  of the result we obtain,  \\[.5cm]
$\mathrm{arg[Tr(\hat{\rho}_1\hat{\rho}_2\hat{\rho}_3)]}$
 $$\mathrm{= q_{01}p^\prime_{01}-q^\prime_{01}p_{01}+q_{02}p^\prime_{02}-q^\prime_{02}p_{02}+q^{\prime}_{01}p^{\prime\prime}_{01}-q^{\prime\prime}_{01}p^{\prime}_{01}+q^{\prime}_{02}p^{\prime\prime}_{02}-q^{\prime\prime}_{02}p^{\prime}_{02}}$$
  $$\mathrm{+\,q^{\prime\prime}_{01}p_{01}-q_{01}p^{\prime\prime}_{01}+q^{\prime\prime}_{02}p_{02}-q_{02}p^{\prime\prime}_{02}+\tan^{-1}\left(\frac{Y_1}{X_1}\right)+\tan^{-1}\left(\frac{Y_2}{X_2}\right)}$$\\
which is the area formed by the phase space points. It is the geometric phase in terms of phase space variables.\\[.5cm]
where  $\mathrm{Y_1=\, q_{01}p^\prime_{01}-q^\prime_{01}p_{01}+ q^{\prime}_{01}p^{\prime\prime}_{01}-q^{\prime\prime}_{01}p^{\prime}_{01}+q^{\prime\prime}_{01}p_{01}-q_{01}p^{\prime\prime}_{01}} $
$$\mathrm{\hspace{2cm}+({q^\prime}^2_{01}+{p^\prime}^2_{01})(q_{01}p^{\prime\prime}_{01}-q^{\prime\prime}_{01}p_{01})+({q^{\prime\prime}}^2_{01}+{p^{\prime\prime}}^2_{01})(q^\prime_{01}p_{01}-q_{01}p^\prime_{01})}$$
$$\mathrm{+(q^2_{01}+p^2_{01})(q^{\prime\prime}_{01}p^\prime_{01}-q^\prime_{01}p^{\prime\prime}_{01})}$$

$\mathrm{X_1=\,\{(q^\prime_{01}q^{\prime\prime}_{01}+p^\prime_{01}p^{\prime\prime}_{01})^2+(q^{\prime\prime}_{01}p^\prime_{01}-q^\prime_{01}p^{\prime\prime}_{01})^2+(q^\prime_{01}q^{\prime\prime}_{01}+p^\prime_{01}p^{\prime\prime}_{01})\}(q^2_{01}+p^2_{01})}$
$$\mathrm{+(q^{\prime\prime}_{01}q_{01}+p^{\prime\prime}_{01} p_{01})({q^\prime}^2_{01}+{p^\prime}^2_{01})+(q^{\prime}_{01} q_{01}+ {p^\prime}_{01}p_{01})({q^{\prime\prime}}^2_{01}+{p^{\prime\prime}}^2_{01}) }$$
$$\mathrm{+(q_{01}q^\prime_{01}+q^\prime_{01}q^{\prime\prime}_{01}+q^{\prime\prime}_{01}q_{01})+(p_{01}p^\prime_{01}+p^\prime_{01}p^{\prime\prime}_{01}+p^{\prime\prime}_{01}p_{01})+1}$$\\
Similarly we can calculate $\mathrm{X_2}$ and $\mathrm{Y_2}$. It can also be calculated in terms of angle of polarisations at which the unitary evolutions were performed.\\[1 cm]

\section{Summary and Discussion}
It has been established that the geometric phase is a symplectic area  produced by the elements in $\cal R$ space. This area is expressed in terms of phase space variables which are used to describe the vertices of geodesic triangle formed by the null phase curves (NPC) in ray space $\cal R$. Here the unitary evolution of density operator is made in ray space $\cal R$. One can also make the evolution in the normalized state vector space called $\cal B$ by taking any vertical lift with an inverse map from ray space($\cal R$) to state space ($\cal B$). The distinct feature of the present work is to avoid the fock state representation of the density operator because of the reason already stated. The possible reason for the appearance of this type of phase phase is due to the characteristic behaviour of the underlying space where we are making the evolution. Here we confine our discussion to calculate geometric phase merely for pure states using this approach. A differential geometric approach has already been proposed towards geometric phase for mixed states by S. Chaturvedi et al [10].\\
\newline
\newline

\textbf{Acknowledgement:} S.S would like to thank to the University Grants Commission, India for providing financial support through Project No:(PSW-065/14-15, Date 2nd Feb-15, S.No.223742). P.M. would also like to thank to the DST, Govt. of  India for financial support through fellowship programme. Authors are thankful to Prof. Parthasarathi Majumdar at Ramakrishna Mission Vivekananda University, Howrah, India for some valuable comments.

\end{document}